\documentclass[
    ,final            
  ]
  {aipproc}

\usepackage{bm}

\layoutstyle{6x9}

\newcommand{\be}{\begin{equation}}
\newcommand{\ee}{\end{equation}}
\newcommand{\bea}{\begin{eqnarray}}
\newcommand{\eea}{\end{eqnarray}}
\newcommand{\Gam}{\Gamma}
\newcommand{\gym}{\gamma_{ab}}

\begin{document}

\title{Non-abelian Yang-Mills in Kundt spacetimes}


\author{Andrea Fuster}{
  address={NIKHEF, Kruislaan $409$, $1098$ SJ, Amsterdam}
}

\begin{abstract}
We present new exact solutions of the Einstein-Yang-Mills system. The solutions are described by a null Yang-Mills field in a Kundt spacetime. They generalize a previously known solution for a metric of $pp$ wave type. The solutions are formally of Petrov type III.
\end{abstract}

\maketitle

\section{Introduction}
The Einstein-Yang-Mills (EYM) theory in four dimensions has been extensively studied in various contexts. One has to make the distinction between solutions which are effectively abelian and truly non-abelian ones. In the first case they are just embeddings of Einstein-Maxwell (EM) solutions. A procedure to construct abelian-like solutions of EYM from known EM ones was given in~\cite{Yasskin:1975ag}. On the other hand, the study of non-abelian EYM has been mainly devoted to static ($\equiv$ time independent) particle-like solutions: solitons and hairy black holes (for a review and references see~\cite{Volkov:1998cc}). \\

\noindent
The first non-abelian and non-static Yang-Mills solutions in flat space were given by Coleman in~\cite{Coleman:1977ps}. They were later generalized by G\"{u}ven in~\cite{Gueven:1979mb} to non-abelian EYM solutions where spacetime was described by the $pp$ wave metric. Recently, Coleman's solution was embedded in Petrov type III Kundt spacetimes~\cite{Fuster:2005qt}, which can be regarded as a generalization of $pp$ waves. We will summarize this result here, while a physical interpretation of these solutions remains an open question\footnote{See, however, ~\cite{Fuster:2005qt} for some hints coming from the geodesic equations.}.      
   
\section{Non-abelian Yang-Mills in flat and \bm{$pp$} wave spacetimes}
We first review the results by Coleman and G\"{u}ven. We use light-cone coordinates $u$, $v$ and complex conjugate coordinates $z$, $\bar{z}$ in the transverse plane. Consider a gauge field of the form:
\be
A_u=(\alpha^a \equiv \lambda^a(u)z+\bar{\lambda}^a(u)\bar{z})\;T_a,\;\;\;A_{\mu}=0 \;\;\mbox{for}\;\;\mu\neq u
\ee
Here $a$ is the gauge index and $\lambda^a$ are arbitrary bounded functions of $u$. The Yang-Mills equation is trivially satisfied for such a field:  
\be
\alpha,_{z\bar{z}}^a=0 \label{ymcomplex}
\ee
These solutions are known as non-abelian plane waves. When gravity is incorporated
one has to solve the coupled EYM equations instead:
\bea
R_{\mu\nu}-\frac{1}{2}g_{\mu\nu}R&=&8\pi T_{\mu\nu}^{\mbox{\tiny YM}} \label{einstein} \\
\nabla_{\lambda}F^{\lambda \rho}-[A_{\kappa},F^{\kappa \rho}]&=&0 \label{curvym} \eea
The YM energy-momentum tensor reads:
\be
T_{\mu\nu}^{\mbox{\tiny YM}}=\frac{\gym}{4\pi}(F_{\mu \eta}^{a}F_{\nu}^{b\;\eta}-\frac{1}{4}g_{\mu\nu}F_{\delta \eta}^{a}F^{b\;\delta \eta})
\ee
Here, $\gym$ is the invariant metric of the Lie group and {\small $F_{\mu \eta}^{a}=\nabla_{[\mu}A_{\eta]}^a+f_{\;\;bc}^aA_{\mu}^bA_{\eta}^c$}. G\"{u}ven proposed a more general ansatz for the YM field where the $\alpha^a$ are arbitrary real functions of $(u,z,\bar{z})$. The corresponding curved YM equation reduces to the flat one when spacetime is of the $pp$ wave type:
\be  
ds^2 = 2\; dz\;d\bar{z}-2\;du\;dv-H(u,z,\bar{z})\; du^2  \label{pp}  
\ee  
There are two reasons why this happens. First, the connection coefficients {\small $\Gam_{\lambda \sigma}^{\;\;\;\;\lambda}$} entering the spacetime covariant derivative in (\ref{curvym}) vanish for the $pp$ wave metric. And second, there are no field strength components of the form {\small $F^{u\rho}$} in this geometry. Equation (\ref{ymcomplex}) is trivially solved by any field of the form: 
\be
{\alpha}^a(u,z,\bar{z})= \chi^a(u,z)+\bar{\chi}^a(u,\bar{z}) \label{ymchi}
\ee
{\small $\chi^a$} being an arbitrary complex function. Such a YM field is null~\cite{Tafel:1986tm}. It is not difficult to see that the Einstein equation is satisfied as well. The only non-zero components of the Ricci and YM energy-momentum tensor are {\small $R_{uu}$} and {\small $T_{uu}^{\mbox{\tiny YM}}$} while the Ricci scalar vanishes. Eq. (\ref{einstein}) reads:
\be
H,_{z\bar{z}}=2\gym \;\alpha,_{z}^a\alpha,_{\bar{z}}^b  
\ee
The solution of the coupled system is given by (\ref{ymchi}) and a $pp$ wave spacetime specified by the function:
\be
H(u,z,\bar{z})=f(u,z)+\bar{f}(u,\bar{z})+2\gym \chi^a \bar{\chi}^b 
\ee 
Here $f$ is an arbitrary complex function. Note that the only solution for which the energy-momentum tensor is bounded throughout spacetime occurs when non-abelian plane waves are recovered, {\small $\chi^a=\lambda^a(u)z$}. On the other hand, the solution becomes effectively abelian in the limit {\small $\alpha^a(u,z,\bar{z})=\beta^a\alpha(u,z,\bar{z})$}, where ({\small $\alpha(u,z,\bar{z})$, $ds^2$}) is a solution of EM and the parameters {\small $\beta^a$} are such that {\small $\gym\beta^a\beta^b=1$}. In any other case the solution has a fully non-abelian character.  

\section{Non-abelian Yang-Mills in Kundt spacetimes}
We will extend the previous result to a more general spacetime in Kundt's class. This class of metrics is described by the line element:
\be
ds^2=-2du\;(dv+W\;dz+\overline{W}\;d\bar{z}+H\;du)  
 +2P^{-2}dzd\bar{z},\; \;\;P,_{v}=0 \label{metgk}
\ee 
Here, $P$ and $H$ are real functions; $W$ is complex. They are algebraically special, i.e., of Petrov type II, D, III, N or O. We will consider type III metrics. In this case $P\equiv 1$ can be taken without loss of generality. Type III vacuum solutions are well-known. A marked difference with respect to $pp$ waves is the dependence of the function $H$ on the light-cone coordinate $v$. Further two distinct cases have to be distinguished according to whether the function $W$ entering (\ref{metgk}) depends on $v$ as well or not. In this last case $pp$ waves arise in the type N limit. The less known Kundt waves appear as the type N limit in the {\small $W,_v\neq 0$} case. Details up to here can be found in~\cite{Stephani:2003tm}. \\

\noindent
The motivation to study exact solutions of gravity-coupled theories described by Kundt metrics is the following. It is well-known that $pp$ waves play an important role in supergravity and string theory. The main reason being the vanishing of the corresponding scalar curvature invariants of all orders. Kundt spacetimes of type III are generalizations of $pp$ waves and have been shown to be the most general metrics with that remarkable property~\cite{Pravda:2002us}. They might therefore have interesting applications in supergravity and string theory similar to $pp$ waves. \\ 


\noindent
In what follows we briefly describe the new embedding of the Yang-Mills field (\ref{ymchi}) in Kundt spacetimes of type III. In this geometry the field is null as well. We make the same distinction in terms of the $v$-dependence of the function $W$ as for vacuum solutions. It can be shown that the curved YM equation reduces in both cases to (\ref{ymcomplex}), the reasons being similar to the $pp$ wave case. Again, there are no {\small $F^{u\rho}$} components of the field strength. Respecting the spacetime derivative, in the first case the only non-zero connection coefficients are precisely {\small $\Gam_{\lambda u}^{\;\;\;\;\lambda}$}. In the second case there are additional coefficients {\small $\Gam_{\lambda z}^{\;\;\;\;\lambda}$}, {\small $\Gam_{\lambda \bar{z}}^{\;\;\;\;\lambda}$} which however cancel each other. The YM energy-momentum tensor is seen to be the same one as in the $pp$ wave case. We give below the precise form of the equations and the corresponding solutions. 
   
\subsection{Solutions with {\normalsize \bm{$W,_{v}=0$}}}
\noindent
Consider the spacetime (\ref{metgk}) specified by:
\be
P=1,\;\;\;
W=W(u,\bar{z}),\;\;\;
H=H^0(u,z,\bar{z}) +\frac{1}{2}\left( W,_{\bar{z}}+\overline{W},_z \right)v \label{3}
\ee
Here $W$ is an arbitrary complex function and $H^0$ is real. Eq. (\ref{einstein}) reads:
\be
H,_{z\bar{z}}^0-\mbox{Re}\left( W,_{\bar{z}}^2+WW,_{\bar{z}\bar{z}}+W,_{\bar{z}u} \right)=2\gym \;\alpha,_{z}^a\alpha,_{\bar{z}}^b \nonumber 
\ee
The solution of the coupled system is given by the YM field (\ref{ymchi}) and:
\be
H^0(u,z,\bar{z})=f(u,z)+\bar{f}(u,\bar{z})+2\gym \chi^a \bar{\chi}^b \\  
 +\mbox{Re}\; \{\left( W,_{u}+WW,_{\bar{z}}    \right)z\}  
\ee 
The type N reduction occurs for {\small $\Psi_3=(1/2)W,_{\bar{z}\bar{z}}=0$}. In this limit the generalized $pp$ waves in~\cite{Gueven:1979mb} are recovered.
\subsection{Solutions with {\normalsize \bm{$W,_{v}\neq0$}}}
\noindent
The second class of solutions is characterized by the $v$-dependence of $W$. The functions in (\ref{metgk}) are now given by: 
\be
P=1,\;\;\;
W=W^0(u,z)-\frac{2v}{z+\bar{z}},\;\;\;
H=H^0(u,z,\bar{z}) + \frac{W^0+\overline{W}^0}{z+\bar{z}}v   
 -\frac{v^2}{(z+\bar{z})^2}  \label{3v} 
\ee
In principle $W^0$ is an arbitrary complex function. In order to solve the Einstein equation explicitly we however consider the specific function {\small $W^0=\;g(u)z$}, {\small $g(u)$} being an arbitrary complex function; this is the simplest function for which the solution is of type III. We further limit the YM field to non-abelian plane waves. Eq. (\ref{einstein}) becomes:
\be
(z+\bar{z})\left( \frac{H^0+g\bar{g}z\bar{z}}{z+\bar{z}} \right)_{,_{z\bar{z}}}-g(u)\bar{g}(u)=2\gym\lambda^a(u)\bar{\lambda}^b(u) 
\ee
The complete solution is described by the YM field {\small $\alpha^a \equiv \lambda^a(u)z+\bar{\lambda}^a(u)\bar{z}$} and:
\be
H^0= (f(u,z)+\bar{f}(u,\bar{z}))\;(z+\bar{z})-g\bar{g}z\bar{z}  
+\sigma(u)(z+\bar{z})^2\left\{ \mbox{ln}(z+\bar{z})-1 \right\}  \label{Kdif}
\ee
Here $\sigma$ is the real function {\small $\sigma(u)=2\gym\lambda^a(u)\bar{\lambda}^b(u)+g(u)\bar{g}(u)$}. The corresponding vacuum solution {\small ($\lambda^a\equiv0$)} has to our knowledge not been considered before. The type N reduction occurs for {\small $\Psi_3=\bar{g}/(z+\bar{z})=0$}. In this case generalized Kundt waves arise.

\section{Conclusions}
We present two (classes of) new exact solutions of the four-dimensional EYM system. The solutions are non-static and have a fully non-abelian character. They might be of interest in supergravity and string theory, in the fashion of $pp$ waves (work in progress).

\bibliography{myproc}

\end{document}